\documentclass[%
 aip,%
 cp,%
 amsmath,%
 amssymb,%
 reprint,%
 groupedaddress,
]{revtex4-1}

\usepackage{graphicx}
\usepackage{dcolumn}
\usepackage{bm}

\newcommand{\ben}{\begin{eqnarray}}
\newcommand{\een}{\end{eqnarray}}
\newcommand{\beq}{\begin{equation}}
\newcommand{\eeq}{\end{equation}}

\begin{document}

\preprint{APS/123-QED}

\title{Non-minimally coupled tachyon field with Noether symmetry\\
under the Palatini approach}

\author{Lucas G. Collodel}
 \email{lgcollodel@gmail.com}
\author{Gilberto M. Kremer}
 \email{kremer@fisica.ufpr.br}
\affiliation{
 Departamento de F\'{i}sica, Universidade Federal do Paran\'{a}, 81531-980 Curitiba, Brazil}


\begin{abstract}
A model for a homogeneous, isotropic, flat Universe composed by dark energy and matter is investigated. Dark energy is considered to behave as a tachyon field, which is non-minimally coupled to gravity. The connection is treated as metric independent when varying the action, providing an extra term to the Lagrangian density. The self-interaction potential and coupling are naturally found by imposing a Noether symmetry to the system. We analyze the evolution of the density parameters and we compare the results obtained for the deceleration parameter, luminosity distance and Hubble parameter with those found in literature from observational data.
\end{abstract}

\maketitle

\section{\label{sec:level1}Introduction}
Dark energy has played a central role in cosmological researches ever since 1998 with the discovery of the currently accelerated expansion of our Universe (see e.g.\cite{A,B}). New theories containing dark components, attempt not only to explain this expansion's feature but also how structures are formed through the evolution of anisotropies, the age of the Universe, the flatness problem and so on. Many different approaches to describe the  nature of dark energy have been made within the last fifteen years, among the most popular ones are the cosmological constant, scalar fields, fermionic fields, aether fields and possibly tachyon fields, a special kind of scalar field with its grounds in string theory, but which can be easily generalized within the framework of classical gravity. Tachyon fields have also been considered to be the inflaton in the early stages of our Universe. The papers \cite{C,C1,C2,C3} consider the minimally coupled tachyon field to behave as dark energy, whereas for the inflaton, minimally coupled tachyon fields were studied with a great variety of self-interaction potentials such as power-laws, exponential, hyperbolic functions of the field (e.g.\cite{D,D1,D2,D3,D4,D5,D6,D7}). In the papers \cite{E,E1} the tachyon field was considered to be the responsible for both inflationary period and the currently accelerated expansion. Constraints on the behaviour of the potential were developed in the work \cite{E}, where it is shown that potentials presenting $V(\phi\rightarrow\infty)\rightarrow 0$ should not be considered, since such behaviour would hinder radiation's dominance in the past. Non-minimally coupled tachyon fields to the curvature were analyzed in \cite{G} to explain the inflationary period where the potential was given in the exponential form and the coupling by a power series of the field. In the work\cite{H} it is also described the nature of dark energy with potential and coupling in the form of power-law functions. Also within this context, a derivative coupling was analyzed in the paper.\cite{I}.

All the works listed above constrained the dynamics to the potentials and couplings inserted into the systems in an ad-hoc way. The Noether symmetry offers great advantage in this matter, as one constrains the solutions only to be compatible with symmetries and the functions are obtained naturally. In the works \cite{J,J1,J2,J3,J4,J5,J6,J7,J8,J9,J10,J11,J12,K}, the approach of symmetry was used to construct different models concerning $f(R)$ gravity, scalar field theories, fermionic fields and finally, the latter one treats a non-minimal coupling between the tachyon field and gravity with Noether symmetry.

In this paper we intend to give a Palatini treatment to the symmetrical constrained coupled tachyon field. Although the scientific community seem to have lost interest in the Palatini approach, the idea of considering a metric independent connection beautifully generalizes the model as one makes no restrictions when varying the action with respect to the dynamical variables.
We consider a homogeneous, isotropic flat expanding Universe, described by the Friedmann-Lema\^itre-Robertson-Walker (FLRW) metric and composed of dark energy and pressureless matter (both baryonic and dark). Dark energy is described as a tachyon field, which is coupled to the curvature scalar. We narrow the solutions of our system by imposing a compatibility with the Noether symmetry, that naturally provides which sample of functions is allowed for the self-interaction potential of the field, as well as for the coupling. Also, when varying the action, we consider the connection to be metric independent, in doing so, our point-like Lagrangian is granted an extra term. The model reproduces satisfactorily the recent behaviour of the Universe when compared to observational data.

The metric adopted is the FLRW for the flat space with signature ($+,-,-,-$). The Levi-Civita connection is written with a tilde, $\tilde{\Gamma}^{\lambda}_{\mu\nu}=\left\{^{\lambda}_{\mu\nu}\right\}$ while the independent connection is given without it $\Gamma^{\lambda}_{\mu\nu}$. Natural constants were rescaled to the unity ($8\pi G=c=1$). Throughout the whole paper, derivative in equations are presented as follows: dots represent time derivatives, while $\partial_{q^i}\equiv\frac{\partial}{\partial q^i}$ and $\partial_{\dot{q}^i}\equiv\frac{\partial}{\partial\dot{q}^i}$ stand for partial derivatives with respect to the generalized coordinate $q^i$ and velocity $\dot{q}^i$, respectively. Furthermore, as reminded in section~\ref{level5}, primes denote differentiation with respect to redshift $z$.
\section{Action}
We consider a Universe composed by both (pressureless) ordinary and dark matter and dark energy, which is here described as a tachyon field ($\phi$) that is non-minimally coupled to the curvature. The action is written as
\beq
\label{actionref}
S=\int d^4x\sqrt{-g}\left[F(\phi)R-V(\phi)\sqrt{1-\partial^{\mu}\phi\partial_{\mu}\phi}-\mathcal{L}_m\right].
\eeq
where $F(\phi)$ is the coupling, $V(\phi)$ is the self-interaction potential, $\mathcal{L}_m$ is the matter field's Lagrangian density and $R$ is the scalar curvature (given in terms of the independent connection $\Gamma^{\lambda}_{\mu\nu}$). Under the Palatini approach, the metric and the affine connection are taken initially to be independent dynamical variables, meaning that the action's variation shall be done separately with respect to these two quantities. The variation with respect to the connection yields the well known expression,
\beq
\Gamma^{\rho}_{\mu\nu}=\tilde{\Gamma}^{\rho}_{\mu\nu}+\frac{1}{2F}\left(\delta^{\rho}_{\nu}\partial_{\mu}F+\delta^{\rho}_{\mu}\partial_{\nu}F-g_{\mu\nu}\partial^{\rho}F\right).
\eeq

\section{Equations of Motion}

We rewrite the Lagrangian density on the spatially flat FLRW metric and, after integrating the second order terms by parts, we get the first order point-like Lagrangian density
\ben\nonumber
\label{pll}
\mathcal{L}&=&-6F\dot{a}^2a-6a^2\dot{a}\partial_{\phi}F\dot{\phi}-a^3V\sqrt{1-\dot{\phi}^2}
\\&-&\frac{3a^3}{2F}(\partial_{\phi}F\dot{\phi})^2-\rho^0_m.
\een
where $a$ is the scale factor and $\rho^0_m$ is a constant and represents the current energy density of the matter fields.

Comparing this Lagrangian density with the one obtained by the metric approach \cite{K}, we notice an extra term $\frac{3a^3}{2F}(\partial_{\phi}F\dot{\phi})^2$. The Friedmann equation is then obtained through the energy equation,
\beq
\label{eneq}
E_{\mathcal{L}}=\dot{a}\partial_{\dot{a}}\mathcal{L}+\dot{\phi}\partial_{\dot{\phi}}\mathcal{L}-\mathcal{L}=0,
\eeq
and reads
\beq
\label{friedmann}
H^2=\left(\frac{\dot{a}}{a}\right)^2=\frac{\rho}{6F}.
\eeq
Here $H$ is the  Hubble parameter and $\rho=\rho_{\phi}+\rho_m$ the total energy density. Furthermore,
\beq
\rho_{\phi}=\frac{V}{\sqrt{1-\dot{\phi}^2}}-6H\partial_{\phi}F\dot{\phi}-\frac{3\left(\partial_{\phi}F\dot{\phi}\right)^2}{2F}; \quad \rho_m=\frac{\rho^0_m}{a^3},
\eeq
denote the tachyon field's  and matter's energy density, respectively.
We may now apply the Euler-Lagrange equations for both degrees of freedom $(a,\phi)$. From the scale factor variation, the acceleration equation follows in the form
\beq
\label{acc}
\frac{\ddot{a}}{a}=-\frac{1}{12F}(\rho+3p_{\phi}),
\eeq
where
\ben
p_\phi&=&-V\sqrt{1-\dot{\phi}^2}-\frac{3\left(\partial_{\phi}F\dot{\phi}\right)^2}{2F}+2\partial^2_{\phi}F\dot{\phi}^2\nonumber
\\
&+&2\partial_{\phi}F\ddot{\phi}+4H\partial_{\phi}F\dot{\phi}
\een
is the tachyon field's pressure.
Similarly, the Euler-Lagrange equation is applied to the tachyon field, leading to the Klein-Gordon equation, which reads
\ben\nonumber
\label{kg}
&&\ddot{\phi}\left(\frac{V}{(1-\dot{\phi}^2)^{3/2}}-3\frac{\left(\partial_{\phi}F\right)^2}{F}\right)+\dot{\phi}^2\left(\frac{3}{2}\frac{\left(\partial_{\phi}F\right)^3}{F^2}
\right.
\\\nonumber
&&\left.
-3\frac{\partial_{\phi}F\partial^2_{\phi}F}{F}\right)
+\dot{\phi}\frac{\dot{a}}{a}\left(\frac{3V}{\sqrt{1-\dot{\phi}^2}}-9\frac{\left(\partial_{\phi}F\right)^2}{F}\right)
\\
&&-6\partial_{\phi}F\left(\frac{\ddot{a}}{a}+\frac{\dot{a}^2}{a^2}\right)+\frac{\partial_{\phi}V}{\sqrt{1-\dot{\phi}^2}}=0.
\een
\section{Noether Symmetry}

The coupling and the self-interaction potential have yet to be specified. It is clear that by choosing them, one narrows the solutions for the equations of motions, as the dynamical system becomes more restricted. Such a choice cannot be made arbitrarily, and when setting the functions in an ad-hoc manner, it shall be substantiated by reasonable arguments, often found analyzing observational data. The coupling cannot vary harshly for example, or there would be an over/under production of $^4$He at the time of nucleosynthesis \cite{L,L1,L2}. As for the potential, most of them are motivated on the grounds of quantum field theory. When applying the Noether symmetry on the other hand, one gets the coupling and potential functions naturally, without having to make any extra impositions about them. Of course it does not make the model more general, but it restricts the solutions to those that exhibit symmetry. If a dynamical system is symmetry compatible, there will always be a conserved quantity, also called constant of motion or Noether charge. There is a special class of vector fields that perform a variational symmetry, and for that reason they are often called generators of symmetry (or complete lift). This vector field is written as
\beq
\textbf{X}\equiv\sum^n_i\alpha^i\frac{\partial}{\partial q^i}+\left(\frac{d}{d\lambda}\alpha^i\right)\frac{\partial}{\partial\dot{q}^i},
\eeq
where the coefficients $\alpha^i$ are functions of the the generalized coordinates ($a$, $\phi$), and $\lambda$ is the independent variable, thus representing the time in our system. According to the Noether theorem, if the Lie derivative of the Lagrangian density along \textbf{X} vanishes, the system carries a conserved quantity. Mathematically speaking
\beq
\textbf{X}\mathcal{L}=L_{\textbf{X}}\mathcal{L}=0\rightarrow L_{\Delta}\left<\theta_L,\textbf{X}\right>=0,
\eeq
where $\Delta=d/dt$ is the dynamical vector field and
\beq
\theta_{\mathcal{L}}=\frac{\partial\mathcal{L}}{\partial\dot{q}^j}dq^j
\eeq
is a Cartan one-form defined locally. It is then clear that the constant of movement is giving by the inner product
\beq
\Theta_0\equiv\left<\theta_{\mathcal{L}},\textbf{X}\right>=\alpha^i\frac{\partial\mathcal{L}}{\partial\dot{q}^i}.
\eeq 
Evaluating the lift on our Lagrangian, yields
\ben
&&\alpha\partial_a\mathcal{L}+\beta\partial_{\phi}\mathcal{L}+\left(\dot{a}\partial_a\alpha+\dot{\phi}\partial_{\phi}\alpha\right)\partial_{\dot{a}}\mathcal{L}\nonumber
\\
&&+\left(\dot{a}\partial_a\beta+\dot{\phi}\partial_{\phi}\beta\right)\partial_{\dot{\phi}}\mathcal{L}=0.
\een
The above equation  is a homogeneous polynomial of degree 2 in the generalized velocities, and since it must be identically zero, every coefficient must vanish. This leads to six partial differential equations, namely
\beq
\label{n1}
6\alpha F+6\beta\partial_{\phi}Fa+12Fa\partial_a\alpha+6a^2\partial_{\phi}F\partial_a\beta=0,
\eeq
\ben
\label{n2}
&&\frac{9a^2\alpha \left(\partial_{\phi}F\right)^2}{2F}-\frac{3a^3\left(\partial_{\phi}F\right)^3\beta}{2F^2}+\frac{3a^3\partial_{\phi}F\partial^2_{\phi}F\beta}{F}\nonumber
\\
&&+6\partial_{\phi}Fa^2\partial_{\phi}\alpha+\frac{3a^3\left(\partial_{\phi}F\right)^2}{F}\partial_{\phi}\beta=0,
\een
\ben
\label{n3}
&&4\partial_{\phi}F\alpha a+2\partial^2_{\phi}F\beta a^2+2\partial_{\phi}Fa^2\partial_a\alpha+4Fa\partial_{\phi}\alpha\nonumber
\\
&&+a^3\frac{\left(\partial_{\phi}F\right)^2}{F}\partial_a\beta+2a^2\partial_{\phi}F\partial_{\phi}\beta=0,
\een
\beq
\label{n4}
-3\alpha a^2V-a^3\beta \partial_{\phi}V=0,
\eeq
\beq
\label{n5}
a^3V\partial_{\phi}\beta=0,\qquad a^3V\partial_a\beta=0.
\eeq

From  (\ref{n4}) we infer that $\alpha=-\frac{\beta a\partial_{\phi}V}{3V}$, while from  (\ref{n5}) we conclude that $\beta=\beta_0$ is a constant. Substituting these coefficients in (\ref{n1}), one finds that $V\propto F$. Solving  (\ref{n2}) and (\ref{n3}) for the self-interaction potential and for the coupling, we find
\beq
\label{potcoup}
V(\phi)=V_0e^{k\phi}; \qquad F(\phi)=F_0e^{k\phi},
\eeq
where $k$, $V_0$ and $F_0$ are constants. Note the linear dependence between the potential and the coupling are the same found in \cite{K}. These results replace the former unknown quantities in the Lagrangian, as well as in the equations of motion.

\section{\label{level5}{Numerical Solutions}}

In order to integrate our equations of motion numerically and plot the curves, we shall change the independent variable from time to redshift, this will turn out to be very useful when setting the initial conditions. From the relation $z=1/a-1$, we infer
\ben
\frac{d}{dt}=-H(1+z)\frac{d}{dz},
\een
where the primes stand for differentiation with respect to the redshift $(z)$. With this change, we shall find numerical solutions for the Hubble parameter and the tachyon field $(H,\phi)$. So we make use of the Friedmann equation (\ref{friedmann}) together with the acceleration equation (\ref{acc}), to give rise to one simple relation
\beq
4F_0e^{k\phi}HH'(1+z)=\rho_{\phi}+\rho_m+p_{\phi}.
\eeq
The Klein-Gordon equation~(\ref{kg}) now becomes:
\ben
&&\left(H^2(1+z)^2\phi''+HH'(1+z)^2\phi'\right)\nonumber
\\\nonumber
&&\times\left(\frac{V_0}{\left[1-H^2(1+z)^2\phi'^2\right]^{3/2}}-3F_0k^2\right)
\\
&&-\frac{3}{2}F_0k^3H^2(1+z)^2\phi'^2+\frac{V_0k}{\sqrt{1-H^2(1+z)^2\phi'^2}}\nonumber
\\
&&-H^2(1+z)\left(\frac{3V_0}{\sqrt{1-H^2(1+z)^2\phi'^2}}-9F_0k^2\right)\nonumber
\\
&&-6F_0k\left[2H^2-HH'(1+z)\right]=0.
\een

These two differential equations, together with the initial conditions, will give us the complete behaviour of the components' densities and pressure, as well as of the scale factor. Before choosing the initial conditions, we may rescale the quantities so they become dimensionless as follows:
\ben
H\rightarrow\bar{H}=\frac{H}{\sqrt{\rho^0}},\quad V\rightarrow\bar{V}=\frac{V}{\rho^0} \nonumber,
\\
\phi\rightarrow\bar{\phi}=\frac{\phi}{\sqrt{\rho^0}},\quad k\rightarrow\bar{k}=k\sqrt{\rho^0}.
\een
where $\rho^0=\rho_{\phi}^0+\rho_m^0$ is the total energy density at the present time. Accordingly, we are now searching solutions for the behaviour of the density parameters, defined as $\Omega_{\phi}=\rho_{\phi}/\rho$ for the tachyon field and $\Omega_m=\rho_m/\rho$ for the matter field.

We need now to consider a few facts concerning the initial conditions $\bar{\phi}'(0)$, $\bar{\phi}(0)$ and $\bar{H}(0)$. Firstly, for the tachyon field to exhibit a negative real pressure at present times, it is required that $\dot{\phi}(0)\ll1$, so it seems very reasonable to set $\bar{\phi}'(0)=10^{-3}$. Hence, we have $\rho_{\phi}(0)\sim V(0)$, which is the same as $\Omega_{\phi}(0)=\bar{V}(0)=0.72$. From (\ref{potcoup}) it follows  $\bar{\phi}(0)=\bar{k}^{-1}\ln(0.72/\bar{V}_0)$. Furthermore, the coupling must equal one-half in present time, so $F_0=\bar{V}_0/1.44$, in a way that we have now, only two free parameters, namely $\bar{V}_0$ and $\bar{k}$. Finally, by the Friedmann equation (\ref{friedmann}), we set $\bar{H}(0)=1/\sqrt{3}$. Fixing $\bar{V}_0=1$, we next plot the results for $\bar{k}=0.1$, $\bar{k}=0.05$ and $\bar{k}=0.01$. Stronger couplings will cause instabilities and are not shown here.

\begin{figure}[!ht]
\centering
\includegraphics[scale=0.6]{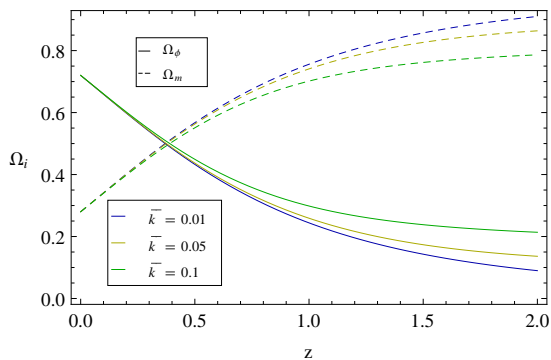}
\caption{Density parameter vs. redshift.}
\label{densityf}
\end{figure}

The evolution of the density parameters are shown in Fig.~\ref{densityf}. The weaker the coupling is, the more rapidly increases the dark energy density while the matter field's decreases. Contrarily, for the strongest coupling $\bar{k}=0.1$, dark energy is present in considerable amount throughout the Universe's evolution for a very long time.
\begin{figure}[!ht]
\centering
\includegraphics[scale=0.6]{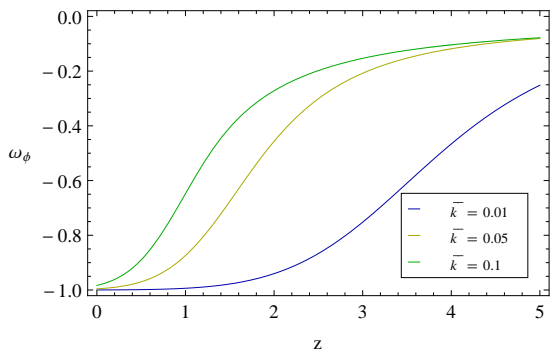}
\caption{$\omega_{\phi}$ vs. Redshift}
\label{ratiof}
\end{figure}

\section{Observational Parameters}
\subsection{Deceleration Parameter}
\begin{figure}[!ht]
\centering
\includegraphics[scale=0.6]{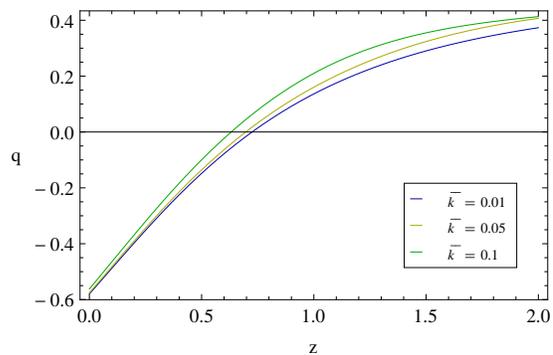}
\caption{Deceleration Parameter vs. Redshift}
\label{dpf}
\end{figure}

The deceleration parameter  -- given by $q=1/2+3p/2\rho$ --  is plotted in Fig.~\ref{dpf}.
As the coupling increases, the transition from decelerated to accelerated expansion happens at lower redshifts. This is already expected and confirmed from Fig.~\ref{ratiof}, where we can clearly see that for stronger couplings, the pressure to energy density ratio, $\omega_{\phi}$, of the tachyon field decreases much more rapidly, assuming values of $\omega_{\phi}>-1/3$ in recent times. In Table~\ref{tabref}, we show some numerical values taken from the curves, where $z_{eq}$ is the redshift  when the densities are equal, $z_t$ stands for the redshift of the decelerated-accelerated transition  and $q(0)$ is the value of the deceleration parameter at present time.

The values found in literature for both $q(0)$ and $z_t$ differ widely from one reference to another, see e.g. \cite{M,M1,M2}. In the work \cite{N}, in order to estimate these values, the authors make 3 different parameterizations containing only 2 free parameters, which are constrained by Supernovae observational data.

\begin{table}[!h]
\begin{ruledtabular}
\begin{tabular}{cccc}
\multicolumn{1}{c}{}& $\bar{k}=0.1$ & $\bar{k}=0.05$ & $\bar{k}=0.01$ \\
\colrule
$z_{eq}$ & 0.3900 & 0.3748 & 0.3702\\
$q(0)$ & -0.5619 & -0.5754 & -0.5798\\
$z_t$ & 0.6316 & 0.6963 & 0.7247\\
\end{tabular}
\end{ruledtabular}
\caption{Values provided by this model}
\label{tabref}
\end{table}

\subsection{Hubble Parameter}

\begin{figure}[!h]
\includegraphics[scale=0.6]{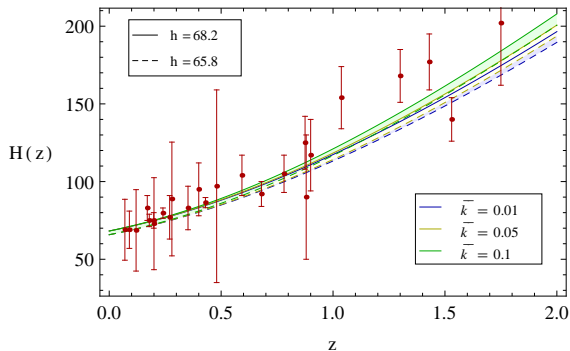}
\caption{Hubble Parameter vs. Redshift}
\label{hpf}
\end{figure}

The Hubble parameter is plotted in Fig.~\ref{hpf} for the three analysed values of $\bar{k}$. The data in red corresponds to observational data from 25 supernovas \cite{hz,hz1,hz2,hz3}. The three curves practically overlap each other for small redshifts but evolve differently as it increases. In future times, more accurate data will provide us enough information to constrain such models.

\section{Final Remarks}
In the beginning of this work, we first wrote the Lagrangian density on the metric and only then varied the system with respect to the generalized coordinates, applying the Euler-Lagrange equations. Point-like Lagrangian is necessary in order to apply the generator of symmetry, but obviously one could derive the same results trading orders, doing the metric variation first, and then writing the equations of motions on the metric. One important result is obtained from the energy-momentum tensor's divergent. The metric variations yields
\ben
\frac{\delta S}{\delta g^{\mu\nu}}&=&FR_{\mu\nu}-\frac{1}{2}g_{\mu\nu}FR +\frac{V}{2}\frac{\partial_{\mu}\phi\partial_{\nu}\phi}{\sqrt{1-\partial^{\mu}\phi\partial_{\mu}\phi}}\nonumber
\\
&+&\frac{1}{2}g_{\mu\nu}\left(V\sqrt{1-\partial^{\mu}\phi\partial_{\mu}\phi}+\frac{\rho^0_m}{a^3}\right)=0,
\een
out of which the component $\mu=\nu=0$ implies the Friedmann equation (\ref{friedmann}). Note that the energy-momentum tensor is not conserved anymore as its four-divergent does not vanish, instead there is an energy flow from the dark energy density to the gravitational field. If we differentiate the Friedmann equation with respect to the time,
\beq
\dot{\rho}+3H(\rho+p)=k\dot{\phi}\rho,
\eeq
and consider that the matter field is not coupled, nor interacts with other fields, the equation above is equivalent to
\beq\label{as}
\dot{\rho_{\phi}}+3H(\rho_{\phi}+p_{\phi})=-k\dot{\phi}\rho_{\phi}.
\eeq

The role of the coupling constant becomes clear at this point. We see that, as already mentioned before, the bigger the constant is, the stronger is the coupling, and from eq. (\ref{as}) we see that this provokes a more intense energy flow from the dark component to the gravitational field. Moreover, on the right hand side of the above equation, we also have a contribution from the generalized velocity of the field and its energy density. However, the tachyon field performs a slow roll, and for all times $\dot{\phi}\ll 1$, not contributing significantly for the energy flow. 

\section{Conclusions}

The present model investigated the dynamics of non-minimally coupled tachyon field, constrained to the Noether symmetry and under the Palatini approach. Although it was not possible to find an analytical solution to the system, the fact that we only had two free parameters (out of which $\bar{V}_0$ did not seem to cause significant difference when admitting a wide range of values), makes it easy to find plausible numerical solutions. The symmetry imposition showed us naturally which functions were allowed to the potential and coupling, and they turned out to be linearly dependent. The coupling is represented by a well behaved function which varies very softly as required, a rapidly varying coupling is not solution to our system. As already mentioned in the previous sections, $\bar{k}$ must be small to avoid instabilities. By inferring that the metric and the connection might be independent from each other, our point-like Lagrangian acquired one extra term. Nevertheless, because both the tachyon field and the coupling function vary smoothly, such term does not contribute quite much for the dynamics. From the solutions, it became clear the importance of the coupling constant, specially as the redshift increases. The comparison with the observational data was very satisfactory as every calculated parameter lies within the observational errors.

\acknowledgments
The authors acknowledge the financial support from CNPq (Brazil).


\nocite{*}
\begin{thebibliography}{9}

\bibitem{M1}
A.~G. Riess et al., \emph{Astrophys. J.} \textbf{607}, 665 (2004).

\bibitem{A}
A.~G. Riess et al., \emph{Astron. J.} \textbf{116},
  1009 (1998).

\bibitem{D}
L.~R. Abramo and F. Finelli, \emph{Phys. Lett. B} \textbf{575},
  165 (2003).

\bibitem{C2}
J.~S. Bagla, and H.~K. Jassal, and T. Padmanabhan, \emph{Phys. Rev. D} \textbf{67},
 063504 (2003).

\bibitem{D7}
L. Balart, and S. del Campo, and R. Herrera, and P. Labra\~{n}a, and J. Saavedra, \emph{Phys. Lett. B} \textbf{647},
 313 (2007).

\bibitem{E1}
V.~H. C\'{a}rdenas, \emph{Phys. Rev. D} \textbf{73},
 103512 (2006).

\bibitem{N}
S. del Campo, and I. Duran, and R. Herrera, and D. Pav\'{o}n, \emph{Phys. Rev. D} \textbf{86},
 083509 (2012).

\bibitem{D4}
C. Campuzano, and S. del Campo, and R. Herrera, \emph{Phys. Lett. B} \textbf{633},
 149 (2006).

\bibitem{J}
S. Capozziello, and A. De Felice, \emph{J. Cosmol. Astropart. Phys.} \textbf{08},
 016 (2008).

\bibitem{J1}
S. Capozziello, and G. Lambiase, \emph{Gen. Relat. Gravit.} \textbf{32},
 295 (2000).

\bibitem{J8}
S. Capozziello, and G. Lambiase, \emph{Grav. Cosmol.} \textbf{6},
 164 (2000).

\bibitem{J7}
S. Capozziello, and S. Nesseris, and L. Perivolaropoulos, \emph{J. Cosmol. Astropart. Phys.} \textbf{12},
 009 (2007).

\bibitem{J6}
S. Capozziello, and R. de Ritis, \emph{Classical Quant. Grav.} \textbf{11},
 107 (1994).

\bibitem{C7}
E.J. Copeland, and M. Sami, and S. Tsujikawa, \emph{Int. J. Mod. Phys. D} \textbf{15},
 1753 (2006).

\bibitem{J5}
M. Demia\ifmmode \acute{n}\else \'{n}\fi{}ski, and R. de Ritis, and G. Marmo, and G. Platania, and C. Rubano, and P. Scudellaro, and C. Stornaiolo, \emph{Phys. Rev. D} \textbf{44},
 3136 (1991).

\bibitem{L2}
F.G. Alvarenga, J.C. Fabris, S.V.B. Gonçalves, and J.A.O. Marinho, \emph{Braz. J. Phys.} \textbf{31},
 546 (2001).

\bibitem{hz3}
E. Gazta\~{n}aga, and A. Cabr\'{e}, and L. Hui, \emph{Mon. Not. R. Astron. Soc.} \textbf{399},
 1663 (2009).

\bibitem{C1}
J.-G. Hao, and X.-Z. Li, \emph{Phys. Rev. D} \textbf{66},
 087301 (2002).

\bibitem{D5}
R. Herrera, and S. del Campo, and C. Campuzano, \emph{J. Cosmol. Astropart. Phys.} \textbf{10},
 009 (2006).

\bibitem{C3}
H.~K. Jassal, \emph{Pramana} \textbf{62},
 757 (2004).

\bibitem{D2}
G.~M. Kremer, and S.~M. Alves, \emph{Gen. Relat. Gravit.} \textbf{36},
 2039 (2004).

\bibitem{M2}
S. Kumar, \emph{Mon. Not. R. Astron. Soc.} \textbf{422},
 2532 (2012).

\bibitem{D1}
D.-J. Liu, and X.-Z. Li, \emph{Phys. Rev. D} \textbf{70},
 123504 (2004).

\bibitem{J11}
M. Demianski, and E. Piedipalumbo, and C. Rubano, and C. Tortora, \emph{Astron. Astrophys.} \textbf{454},
 55 (2006).

\bibitem{hz2}
M. Moresco et al., \emph{J. Cosmol. Astropart. Phys.} \textbf{08},
 006 (2012).

\bibitem{C}
T. Padmanabhan, \emph{Phys. Rev. D} \textbf{66},
 021301 (2002).

\bibitem{G}
Y.-S. Piao, and Q.-G. Huang, and X. Zhang, and Y.-Z. Zhang, \emph{Phys. Lett. B} \textbf{570},
 1 (2003).

\bibitem{C9}
J. Ren, and X.-H. Meng, \emph{Int. J. Mod. Phys. D} \textbf{17},
 2325 (2008).

\bibitem{J4}
R. de Ritis, and G. Marmo, and G. Platania, and C. Rubano, and P. Scudellaro, and C. Stornaiolo, \emph{Phys. Rev. D} \textbf{42},
 1091 (1999).

\bibitem{J9}
S. Capozziello, and P.~K.~S. Dunsby, and E. Piedipalumbo, and C. Rubano, \emph{Astron. Astrophys.} \textbf{51},
 472 (2007).

\bibitem{B}
S. Perlmutter et al., \emph{Astrophys. J.} \textbf{2},
 565 (1999).

\bibitem{E}
M. Sami, and P. Chingangbam, and T. Qureshi, \emph{Phys. Rev. D} \textbf{66},
 043530 (2002).

\bibitem{L}
D.~I. Santiago, and D. Kalligas, and R.~V. Wagoner, \emph{Phys. Rev. D} \textbf{56},
 7627 (1997).

\bibitem{I}
V.~K. Shchigolev, and M.~P. Rotova, \emph{Grav. Cosmol.} \textbf{18},
 88 (2012).

\bibitem{hz}
J. Simon, and L. Verde, and R. Jimenez, \emph{Phys. Rev. D} \textbf{71},
 123001 (2005).

\bibitem{J2}
J.~C.~C de Souza,  and V. Faraoni, \emph{Classical Quant. Grav.} \textbf{24},
 3637 (2007).

\bibitem{K}
R.~C. de Souza, and G.~M. Kremer, \emph{Classical Quant. Grav.} \textbf{26},
 135008 (2009).

\bibitem{J12}
R.~C. de Souza, and G.~M. Kremer, \emph{Classical Quant. Grav.} \textbf{25},
 225006 (2008).

\bibitem{D3}
D.~A. Steer, and F. Vernizzi, \emph{Phys. Rev. D} \textbf{70},
 043527 (2004).

\bibitem{hz1}
D. Stern,  and R. Jimenez, and L. Verde, and M. Kamionkowski, and S.~A. Stanford, \emph{J. Cosmol. Astropart. Phys.} \textbf{02},
 008 (2010).

\bibitem{J3}
B. Vakili, \emph{Phys. Lett. B} \textbf{664},
 16 (2008).

\bibitem{J10}
B. Vakili, and F. Khazaie, \emph{Classical Quant. Grav.} \textbf{29},
 035015 (2012).

\bibitem{M}
J.-M. Virey, and P. Taxil, and A. Tilquin, and A. Ealet, and C. Tao, and D. Fouchez, \emph{Phys. Rev. D} \textbf{72},
 061302 (2005).

\bibitem{D6}
H.-H. Xiong, and J.-Y. Zhu, \emph{Phys. Rev. D} \textbf{75},
 084023 (2007).

\bibitem{L1}
X. Chen, preprint arXiv:0011533v1 [astro-ph] (2000).

\bibitem{hz4}
C. Zhang, and H. Zhang, and S. Yuan, and T.-J. Zhang, and Y.-C. Sun, preprint arXiv:1207.4541v3 [astro-ph.CO] (2014).

\bibitem{C8}
G. Panotopoulos, preprint arXiv:0606249v2 [astro-ph] (2006).

\bibitem{H}
S.~K. Srivastava, preprint arXiv:0409074v4 [gr-qc] (2005).

\end{thebibliography}
\end{document}